\documentclass[lettersize,journal]{IEEEtran}
\usepackage{amsmath,amsfonts}
\usepackage{algorithmic}
\usepackage{algorithm}
\usepackage{array}
\usepackage[caption=false,font=normalsize,labelfont=sf,textfont=sf]{subfig}
\usepackage{textcomp}
\usepackage{stfloats}
\usepackage{url}
\usepackage{verbatim}
\usepackage{graphicx}
\usepackage{cite}
\usepackage{booktabs}
\usepackage{xspace}
\newcommand{\name}{Urania\xspace}
\newcommand{\modelname}{linking network\xspace}

\newcommand{\etal}{{\textit{et~al.}}\xspace}
\begin{document}
\setlength{\abovedisplayskip}{1pt}
\setlength{\belowdisplayskip}{1pt}

\title{Urania: Visualizing Data Analysis Pipelines for Natural Language-Based Data Exploration}

\author{ Yi Guo, Nan Cao, Xiaoyu Qi, Haoyang Li, Danqing Shi, 
  Jing Zhang, Qing Chen, and 
  Daniel Weiskopf
\thanks{ Yi Guo, Nan Cao, and Xiaoyu Qi are with Intelligent Big Data Visualization Lab, Tongji University. E-mail: \{2010937, nan.cao, 2231952\}@tongji.edu.cn. \\
Nan Cao is the corresponding author.}
\thanks{Danqing Shi is with Finnish Center for Artificial Intelligence. E-mail: \{ danqing.shi@aalto.fi\}; Haoyang Li and Jing Zhang are with Renmin University of China. E-mail: \{lihaoyang.cs, zhang-jing\}@ruc.edu.cn.; Daniel Weiskopf is with University of Stuttgart. E-mail: \{Daniel.Weiskopf@visus.uni-stuttgart.de\}}
}



\IEEEaftertitletext{\vspace{-2em}}
\markboth{Journal of \LaTeX\ Class Files,~Vol.~14, No.~8, August~2021}%
{Shell \MakeLowercase{\textit{et al.}}: A Sample Article Using IEEEtran.cls for IEEE Journals}


\maketitle


\begin{abstract}
Exploratory Data Analysis (EDA) is an essential yet tedious process for examining a new dataset. To facilitate it, natural language interfaces (NLIs) can help people intuitively explore the dataset via data-oriented questions. 
However, existing NLIs primarily focus on providing accurate answers to questions, with few offering explanations or presentations of the data analysis pipeline used to uncover the answer. 
Such presentations are crucial for EDA as they enhance the interpretability and reliability of the answer, while also helping users understand the analysis process and derive insights. To fill this gap, we introduce \name, a natural language interactive system that is able to visualize the data analysis pipelines used to resolve input questions. It integrates a natural language interface that allows users to explore data via questions, and a novel data-aware question decomposition algorithm that resolves each input question into a data analysis pipeline. This pipeline is visualized in the form of a datamation, with animated presentations of analysis operations and their corresponding data changes. Through two quantitative experiments and expert interviews, we demonstrated that our data-aware question decomposition algorithm outperforms the state-of-the-art technique in terms of execution accuracy, and that \name can help people explore datasets better. In the end, we discuss the observations from the studies and the potential~future~works.

\end{abstract}

\begin{IEEEkeywords}
Natural Language Interface
\end{IEEEkeywords}

\vspace{-1em}
\section{Introduction}
Exploratory Data Analysis (EDA) is an important analytics process for knowledge discovery~\cite{milo2020automating} in which analysts interactively explore data by performing a series of analysis operations (e.g., filter, aggregate). 
Although effective, performing EDA requires a great deal of time and profound analytical skills~\cite{milo2020automating}. Thus, a variety of techniques have been devised over the last two decades to facilitate EDA. Commercial software, such as Excel~\cite{excel} and Tableau~\cite{tableau}, have been proposed to help users explore data without programming. To further save users' efforts, automatic EDA techniques have been developed to recommend meaningful insights within data~\cite{shi2020calliope,vartak2015seedb,bar2020automatically} or high-utility operations to be performed in the next exploratory step~\cite{milo2018next,somech2019predicting}. In recent years, a surge of natural language interfaces (NLIs)~\cite{dhamdhere2017analyza,yu2019flowsense,srinivasan2021snowy,shi2021talk2data} have been introduced to help users explore the data more intuitively. Users can directly ``ask" the analysis system for the answers to their questions, making them focus on the analysis task rather than the manipulation of the systems~\cite{shen2021towards,hoque2017applying}. 

\vspace{0.25mm}
However, existing NLIs mostly focus on better understanding questions and delivering more precise answers. The data analysis pipelines that uncover the answers are not explained or presented in existing NLIs. Such presentations are important for EDA, as they enhance the interpretability and trustworthiness of the answers while helping users understand the analysis process and derive insights~\cite{pu2021datamations,deutch2020explained,deutch2022fedex}. Yet, to visualize the data analysis pipeline in NLIs, there are three challenges that need to be addressed. 
First, data analysis pipelines cover multiple data dimensions and incorporate a series of analysis operations, which is difficult for a system to construct. Second, data analysis pipelines usually cannot be properly organized without taking into account both questions and underlying datasets. However, it remains a challenge for machines to comprehend the information within the datasets and relate them to the questions~\cite{shen2021towards,wang-etal-2020-rat}.
Third, to better present data analysis pipelines, intermediate operations and their corresponding data states should be arranged in order and visualized in a form that facilitates~narration~and~interpretation.

To address the aforementioned challenges, we introduce \name, 
a natural language-based interactive system that is able to resolve data-related questions and visualize the data analysis pipelines used to uncover the answers.
Specifically, when a user inputs a dataset and a question, \name first utilizes a data-aware question decomposition algorithm to break down the question into a data analysis pipeline that reveals the answer. 
After that, the generated pipeline drives an action-oriented unit visualization to
create a datamation~\cite{pu2021datamations}, which is an animated presentation that showcases the analysis operations within the pipeline and their corresponding data changes.
Here, we adopt the term ``unit visualization" from Drucker and Fernandez~\cite{drucker2015unifying}, describing the class of visualization techniques that has a direct mapping between individual data items and corresponding unique visual marks. Finally, the generated datamation is displayed in our elaborately designed user interface, where users can refine it or further explore the dataset along it using rich interactions. To validate our work, we conducted two quantitative experiments for evaluating the performance of the data-aware question decomposition algorithm.  Additionally, we interviewed three domain experts to evaluate the usability of \name by comparing it with Tableau~Ask~Data~\cite{tableau}.
\vspace{1mm}

In summary, the contributions of the paper are as follows:
\begin{itemize}

\item {\bf Algorithm.} We introduce a data-aware question decomposition algorithm that is able to incorporate the data--question links in the question decomposition process to generate more accurate data analysis pipelines. The algorithm employs a \textit{\modelname} built upon the language model RoBERTa~\cite{liu2019roberta} to understand the input dataset, identify data--question links, and inject these relations into a question \textit{decomposition network} to make the generated data analysis pipeline~more~accurate. 



\item {\bf System.} We developed \name, an interactive system that integrates an NLI and the aforementioned question decomposition algorithm to visualize the data analysis pipeline for each input question. With the system, users are provided with datamations~\cite{pu2021datamations} as answers to their questions and are allowed to interact with the datamations to choose better analysis steps or fix errors.

\item{\bf Evaluation.} We conducted two quantitative experiments to validate the performance of the data-aware question decomposition algorithm, and an interview and case study with three expert users to demonstrate the usability of~the~\name. 

\end{itemize}

\section{Related Work}
In this section, we review the existing studies that are most relevant to our work: question decomposition, probing the data analysis pipeline, and NLIs for data visualization.

\subsection{Question Decomposition}
Question decomposition is an active topic in natural language processing that decomposes a complex question into easier sub-questions or a reasoning process. Over the last decade, a variety of works have been devised to decompose the questions that are related to different modalities, such as question answering based on the image~\cite{andreas-etal-2016-learning,johnson2017inferring}, knowledge graph~\cite{hu2021edg,bhutani2019learning}, documentation~\cite{perez-etal-2020-unsupervised,min-etal-2019-multi}, and structured~database~\cite{saparina-osokin-2021-sparqling,wolfson-etal-2022-weakly}. 

To decompose the questions related to image, Johnson~\etal~\cite{johnson2017inferring} proposed a model that consists of a generator to construct reasoning programs and an execution engine to execute these programs in the context of images to produce answers.
Similarly, 
the idea of ``decompose--execute--join'' also extends to the knowledge-based question-answering techniques. For instance, Zheng~\etal~\cite{zheng2018question} decomposed a complex question into multiple simple questions via templates, where each question is parsed into a simple logic form. Next, intermediate answers are generated via these simple logic forms, and final answers are jointly obtained. 

The works of Saparina~\etal~\cite{saparina-osokin-2021-sparqling} and Wolfson~\etal~\cite{wolfson-etal-2022-weakly} translate data questions into SQL or SPARQL queries with QDMR~\cite{wolfson2020break} as intermediate representations. Wolfson~\etal~\cite{wolfson-etal-2022-weakly} fine-tunes the pre-trained language model T-5~\cite{wolf2019huggingface}
to translate the question into a QDMR sequence, and Saparina~\etal~\cite{saparina-osokin-2021-sparqling} build an encoder--decoder-based neural model with a grammar-based decoder~\cite{yin-neubig-2017-syntactic} to implement such translation.
Yet, both papers suffer from low precision, yielding a variety of errors in the decomposed results.  To address this problem, we introduce a novel data-aware question decomposition algorithm to identify data--question links and incorporate them into the question decomposition process. Quantitative experiments show our algorithm can precisely identify question-data relations and generate more accurate QDMR sequences~compared~to~\cite{wolfson-etal-2022-weakly,saparina-osokin-2021-sparqling}.

\vspace{-1em}
\subsection{Probing the Data Analysis Pipeline}

Data insights are sensitive to the analytics procedures used to uncover them. A minor change in the procedure can lead to a vastly different result~\cite{herndon2014does}. Therefore, the EDA notebook~\cite{kery2018story,rule2018exploration,bar2020automatically} becomes an increasingly popular method for data analysts to create illustrative exploratory programs and share data insights on online platforms. Analysts can document the steps they took to derive their insights in EDA notebooks, making the insights more accessible and actionable for a wider audience. Although effective, if the notebook is not well documented or the analysis programs are not accompanied by clear explanations, it is hard for viewers to follow and understand what exactly is important in each exploratory step. This issue can be solved by  automatically attaching explanations for EDA notebooks \cite{deutch2020explained,deutch2022fedex}. These techniques inspect the data state at each analysis step and derive interesting insights from it. The insights are transformed into explanations in the form of natural language captions or visualizations. 

The recent work by Pu~\etal~\cite{pu2021datamations} introduces the idea of datamation, which is the animation designed to interpret the complex data analysis pipeline step by step. 
Datamation enhances a static visualization with details from the data analysis phase to convey important insights and help people understand specific analysis results in everyday settings. Following this idea, we present the underlying data analysis pipelines for user questions in form of datamations, enhancing the interpretability and credibility of the answers.

\vspace{-1em}
\subsection{Natural Language Interface for Data Visualization}
NLIs for data visualization emerged as a promising way of interacting with data and performing analytics via natural language. It has drawn great attention from research community~\cite{setlur2016eviza,narechania2020nl4dv,hoque2017applying,srinivasan2020interweaving, shi2021talk2data, luo2021natural, yu2019flowsense, wang2022towards, liu2021advisor} and industry~\cite{powerbi,tableau}. 

There has been a good improvement in the ability to resolve questions, however, most of the NLIs still lack an interpretation of how the answer was derived. Feng~\etal~\cite{feng2023xnli} were aware of this issue and introduced XNLI, an NLI that elucidates the underlying query interpretation process to help users understand the capability of NLI and fix problems within their queries.
In contrast to XNLI, the focus of \name is to explain the analysis process in response to user inquiries, rather than the mechanism of interpreting the~queries.

Another recent work that shares similarities with ours is DataParticles~\cite{cao2023dataparticles}, an NLI for authoring data stories with animated unit visualizations. DataParticles enables users to provide textual scripts of the story, which are then translated into appropriate visualizations and animations. Different from our system, DataParticles is primarily an authoring tool designed to create data stories, rather than explore the data. Additionally, DataParticles employs a basic constituency parsing technique to comprehend the input text, which lacks the capability to generate a sequence of analytical operations for extracting the answer to the~given~question.

In this work, we introduce \name, an NLI utilized to facilitate EDA. It is able to generate datamations as the answers to input questions, enhancing the interpretability of the results and helping users explore data more productively.

\vspace{-1em}
\section{System Overview}
In this section, we describe the design requirements of the \name system, followed by an introduction to the system's architecture and the definition of QDMR.

\vspace{-1em}
\subsection{Design Requirements}
\name system has been designed to fulfill several real-world requirements for facilitating exploratory data analysis. The concrete requirements were derived based on a 
combination of reviewing prior NLIs and discussions with three domain experts. As a result, four requirements were derived. 

\begin{enumerate}

\item[{\bf R1}] {\bf Supporting natural language questions.} The system should support data questions in natural language as input to help users focus on exploring their data rather than manipulating the system~\cite{shen2021towards,hoque2017applying}. 

\item[{\bf R2}] {\bf Visualizing data analysis pipelines.} 
Insights are sensitive to the methods used to obtain them. Minor changes in analysis steps can lead to vastly different outcomes~\cite{herndon2014does}. The system should not only be able to find the correct answer for the input question, but more importantly, it needs to visualize the data analysis pipeline to interpret the answer~\cite{pu2021datamations,rule2018exploration}.

\item[{\bf R3}] {\bf Creating comprehensible presentations.} The system needs to create comprehensible presentations of data analysis pipelines that are easy to follow and understand. The presentations should illustrate the changes between any two succeeding steps in the analysis pipeline without losing focus or increasing~cognitive~load~\cite{woods1984visual}.

\item[{\bf R4}] {\bf Enabling modification of generated pipelines.} The system should provide an interface for users to interact with the generated data analysis pipelines. This feature enables users to fix the mistakes from the algorithm and delve deeper~into~the~data.
\end{enumerate}
\vspace{-1em}
\subsection{System Architecture and Running Pipeline}    
\begin{figure}[!tbh]
\setlength{\abovecaptionskip}{10pt}
\centering 
\includegraphics[width=0.9\columnwidth]{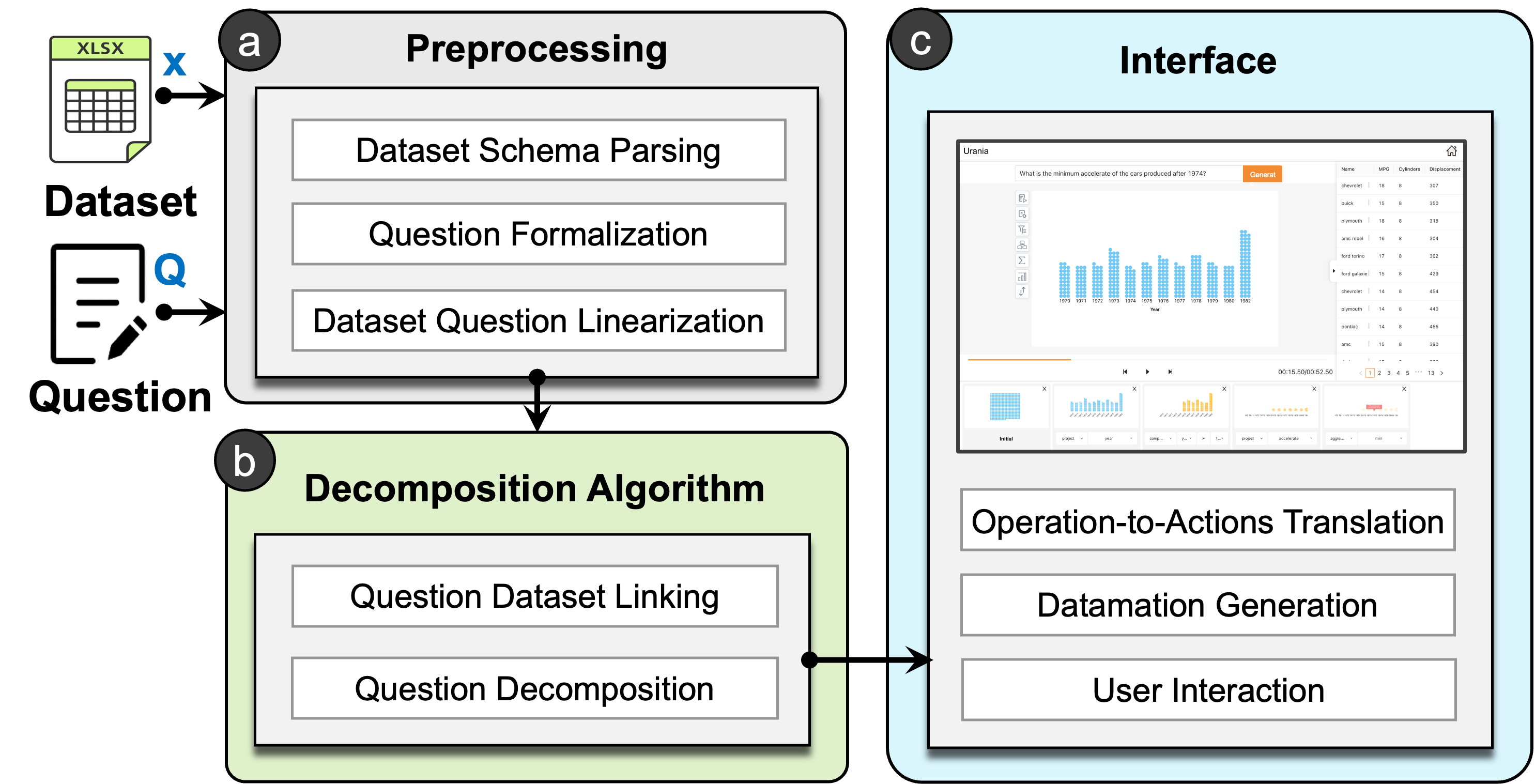}
\vspace{-1em}
\caption{The architecture of \name system consists of three major modules: (a)~preprocessing module, (b)~decomposition algorithm, (c)~NL interface.
}
\vspace{-0.5em}
\label{fig:pipeline}
\end{figure}

To fulfill the above requirements, we introduce \name system. As shown in Fig.~\ref{fig:pipeline}, it consists of three major modules: (a) the \textit{Preprocessing Module}, (b) the \textit{Decomposition Algorithm}, and (c) the \textit{Natural Language Interface}. Specifically, when a user uploads a tabular dataset $X$ and inputs a data question $q_x$ in natural language ({\bf R1}), the \textit{Preprocessing Module} (Fig.~\ref{fig:pipeline}(a)) initially
extracts the schema $s$ of the dataset, which identifies the tables, columns, and structure of $X$. The $s$ is subsequently converted to a word sequence $s_x$ with special tokens inserted in order to facilitate the calculations in the following steps.

Next, the \textit{Decomposition Algorithm}~(Fig.~\ref{fig:pipeline}(b)) resolves $q_x$ in the context of a data schema $s_x$ and decomposes it into a data analysis pipeline represented as a sequence of QDMR operations~\cite{wolfson2020break}:
\begin{equation}
\label{eq:decompose}
S_{op} = [op_{1}, op_{2}, ..., op_{n}] \leftarrow Decompose ({q_x, s_x})
\end{equation}
where each operation $op_i$ indicates an analytic calculation on the underlying dataset. Performing these operations in order will uncover the answer to the $q_x$, and $S_{op}$ represents~a~data~analysis~pipeline. 
 
Finally, the $S_{op}$ is visualized in the form of a datamation ({\bf R2}), with the animated presentation of each operation and its corresponding data changes ({\bf R3}). The datamation is displayed in a \textit{Natural Language Interface}~(Fig.~\ref{fig:pipeline}(c)) that users are allowed to edit~(i.e., add, delete, modify) the operations in $S_{op}$ ({\bf R4}) or input a new question. Once operations are edited, the datamation is updated accordingly. 
 
\vspace{-1em}
\subsection{QDMR}
\label{sec:qdmr}
\begin{table*}[!tbh]
    \setlength\aboverulesep{0pt}
    \setlength\belowrulesep{0pt}
    \def\arraystretch{1.3}
\begin{tabular}{p{0.1\linewidth}wc{0.2\linewidth}p{0.3\linewidth}wc{0.075\linewidth}wc{0.075\linewidth}wc{0.075\linewidth}}
\toprule

\multicolumn{3}{c|}{\textbf{(a) QDMR OPERATIONS}}           & \multicolumn{3}{c}{\textbf{(b) UNITVIS ACTIONS}} \\\midrule
\textbf{Operation} &
\multicolumn{1}{c}{\textbf{Arguments}}    &
 \multicolumn{1}{c|}{\textbf{Description}}  &
 \multicolumn{1}{c}{\textbf{Data}} &
 \multicolumn{1}{c}{\textbf{Visual}}  &
 \multicolumn{1}{c}{\textbf{Annotation}} \\\midrule
SELECT &  \begin{tabular}[c]{@{}c@{}}
table/column
\end{tabular}   &
\multicolumn{1}{l|}{\begin{tabular}[c]{@{}l@{}}
Select data records from a data source \\ (i.e., table/column)
 \end{tabular}}
& select & layout & ---
\\\hline

PROJECT &  column     
& \multicolumn{1}{l|}{Retrieve the \textit{column} values from records }  & --- & 
\begin{tabular}[c]{@{}c@{}}
size [numerical] \\ color [categorical] \\x-axis [temp./categ.]
\\y-axis [temp./categ.]
\end{tabular}

& ---
\\ \hline
FILTER  & 
\begin{tabular}[c]{@{}c@{}}
condition
\end{tabular}
& 
\multicolumn{1}{l|}{
\begin{tabular}[c]{@{}l@{}}
Filter the data based on a condition \\(i.e., $=, \neq, >, <, \ge, \le$) \end{tabular} }
 & filter & --- & highlight, hide
 \\\hline

SUPERLATIVE  & 
\begin{tabular}[c]{@{}c@{}}
column, \\superlative condition  
\end{tabular}
& \multicolumn{1}{l|}{\begin{tabular}[c]{@{}l@{}}Find a record whose  \textit{column} has the \\ \textit{maximum/minimum} value\end{tabular}}  & filter & --- & highlight, hide                      
\\ \hline
AGGREGATE    &  agg methods                  &  
\multicolumn{1}{l|}{\begin{tabular}[c]{@{}l@{}} Compute the \textit{max/min/sum/count/avg} value
\end{tabular}} & aggregate & --- & annotate                             
\\ \hline
GROUP        & 
\begin{tabular}[c]{@{}c@{}}
column \\ agg methods 
\end{tabular}  
& \multicolumn{1}{l|}{\begin{tabular}[c]{@{}l@{}}
Group data by the \textit{column} and compute the \\ \textit{max/min/sum/count/avg} value of each group
\end{tabular}}  & --- & 
\begin{tabular}[c]{@{}c@{}}
x-axis [temporal]\\
y-axis [categorical]
\end{tabular}
& annotate
\\ \hline
SORT         & 
\begin{tabular}[c]{@{}c@{}}
 attribute,  \textit{asc/desc}
\end{tabular} 
& \multicolumn{1}{l|}{\begin{tabular}[c]{@{}l@{}}Sort the data by \textit{column} in a \textit{asc/desc} order
\end{tabular}}   & sort & --- & ---   
 \\\bottomrule          
\end{tabular}
\vspace{2mm}
\caption{The QDMR operations and the corresponding actions designed to drive a dynamic unit visualization.}
\vspace{-3em}
\label{tab:operatorstable}

\end{table*}

QDMR~\cite{wolfson2020break}, Question Decomposition Meaning Representation, is an approach of representing a question by a sequence of operations that can be executed to answer the question. Each operation is responsible for querying the source data or analyzing the outputs of the previous operation(s). It is a data-independent representation that can be applied to many NLP benchmarks. Wolfson~\etal~\cite{wolfson2020break} released Break, a QDMR dataset with 13 types of operations and 83,978 questions over ten NLP benchmarks, such as data-related questions from Spider~\cite{yu-etal-2018-spider}, document-related questions from HotPotQA~\cite{yang-etal-2018-hotpotqa}, and image-related questions from CLEVR~\cite{johnson2017clevr}. In our work, we focus on seven types of operations employed to represent the data-related questions in Spider~\cite{yu-etal-2018-spider}.
The formalization and detailed explanation of each operation are provided in Table~\ref{tab:operatorstable}(a).

\begin{figure}[!tbh]
\centering 
\vspace{-1em}
\includegraphics[width=0.9\columnwidth]{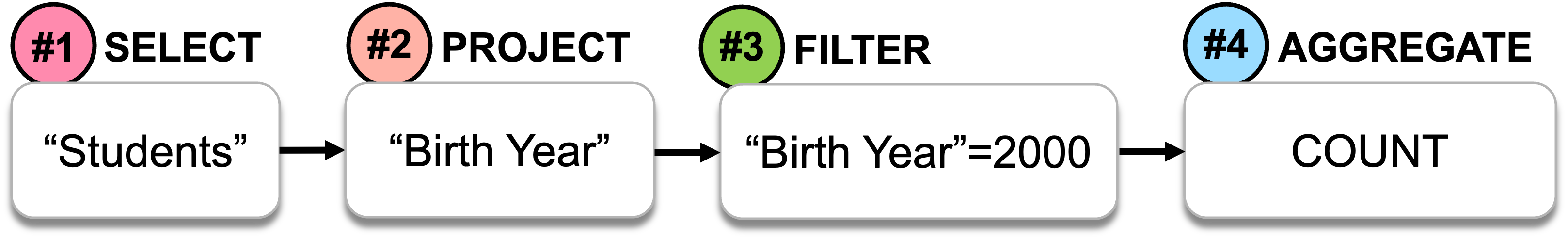}
\caption{QDMR operations chain to answer the question: ``how many students were born in 2000?''}
\label{fig:qdmr}
\end{figure}

Fig.~\ref{fig:qdmr} demonstrates an example of using QDMR to resolve the question: \textit{how many students were born in 2000?} A sequence of four operations is used to answer this question:
\begin{itemize}
    \item Step 1: \texttt{SELECT} (``Students"). \texttt{SELECT} is an operation used to retrieve records from the data. It is similar to the ``select" statement in the structured query language (SQL). The argument \textit{``Students"} corresponds to a column or table name in the dataset. In step 1, the intent of the operation is to select all the student records from the data for further exploration.
    
    \item Step 2: \texttt{PROJECT} (``Birth Year").  \texttt{PROJECT} is an operation that retrieves a certain attribute from the source/input records. The argument \textit{``Birth Year''} specifies the attribute to be retrieved, which is usually a data column. In step 2, the intent of the operation is to retrieve the year of birth of previously selected students.
    
    \item Step 3: \texttt{FILTER} (``Birth Year''$=$2000). The \texttt{FILTER} operation uses a condition to filter records based on a specified attribute. The argument  \textit{``Birth Year''$=$2000} is the comparative condition based on which the data will be filtered. Step 3 finds out the students born in 2000.
    
    \item Step 4: \texttt{AGGREGATE} (count). \texttt{AGGREGATE} computes the aggregated value of the records. It supports six frequently used aggregation methods: count, max, min, sum, avg, and median. The argument indicates the aggregation method. Step 4 counts the number of students in the filtered records.

\end{itemize}

\vspace{1mm}

\section{Data-Aware Question Decomposition}
\begin{figure}[!tbh]
\centering 
\vspace{-1em}
\includegraphics[width=0.8\columnwidth]{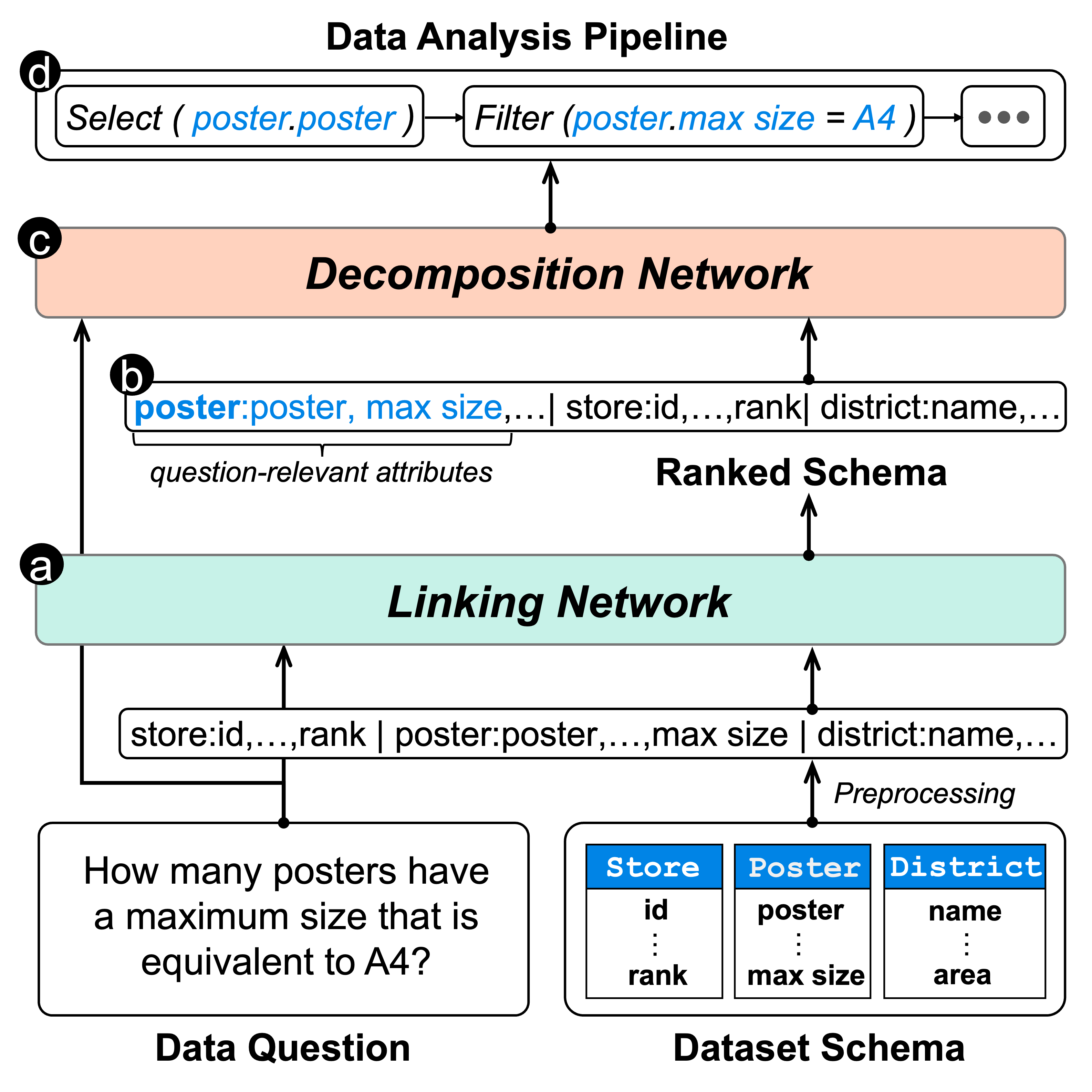}
\caption{The pipeline of Data-Aware Question Decomposition Algorithm.}
\vspace{-1.75em}
\label{fig:calibrate}
\end{figure}

In the system, an algorithm~(Fig.~\ref{fig:calibrate}) has been introduced to incorporate data--question links in the question decomposition process to make the generated data analysis pipelines more accurate. Given an input question and an underlying dataset, the algorithm first employs a \modelname $\mathcal{L}$~(Fig.~\ref{fig:calibrate}(a)) to analyze their semantics and discover the connections between them~(Fig.~\ref{fig:calibrate}(b)). With this information, the algorithm then uses a decomposition network~$\mathcal{D}$~(Fig.~\ref{fig:calibrate}(c)) to break down the input question into an analysis pipeline represented as a sequence of QDMR operations along with their~respective~parameters~(Fig.~\ref{fig:calibrate}(d)). 
\vspace{-1em}
\subsection{Linking Network $\mathcal{L}$}
\label{sec:ranking}

$\mathcal{L}$ is a deep neural network designed to link the question with the underlying dataset, identifying references of columns and tables in the input question. This process is crucial for question decomposition, as
the identified data--question links could help $\mathcal{D}$ learn the conversion between questions and data analysis pipelines, reducing the input's sparsity and enhancing the output's accuracy. 
\begin{figure}[!tbh]
\centering 
\vspace{-1em}
\includegraphics[width=0.8\columnwidth]{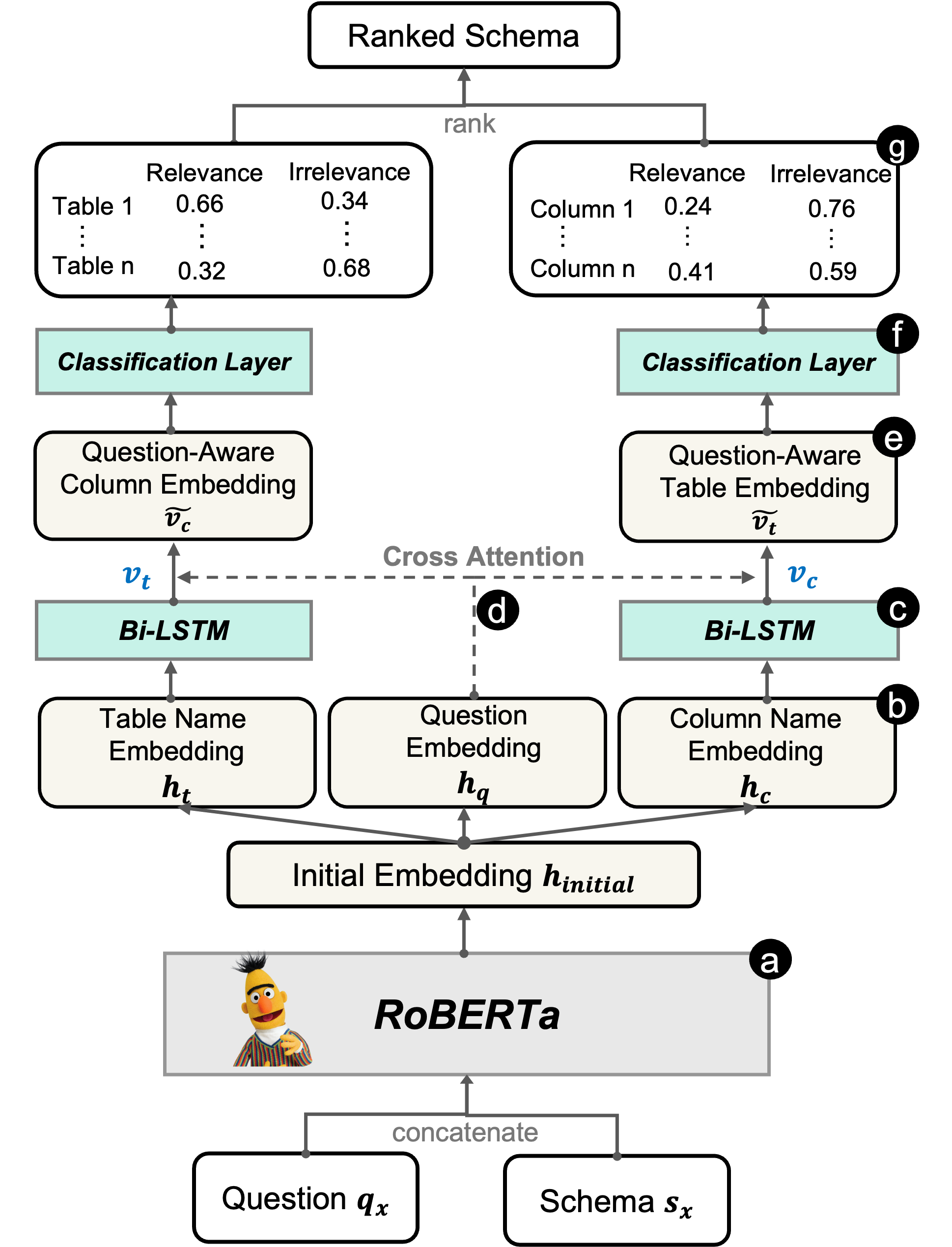}
\caption{The architecture of the \modelname.}
\label{fig:rankingnetwork}
\end{figure}

Fig.~\ref{fig:rankingnetwork} depicts the architecture of $\mathcal{L}$. It takes the question and the dataset schema as input to determine the probabilities of each data attribute~(a table or a column) being associated with the input question. 
Relevant and irrelevant data attributes are categorized based on these probabilities, which explores links between the dataset and the question. The data attributes are then ranked according to their probabilities, as recommended in~\cite{li2023decoupling}, resulting in a ranked schema infused with data--question links, where the order of the attributes reflects their degree of relevance to the question. This schema is then passed on to $\mathcal{D}$ to incorporate the links into the question decomposition process.




We built $\mathcal{L}$ upon the pre-trained language model RoBERTa~\cite{liu2019roberta} (Fig.~\ref{fig:rankingnetwork}(a)), a state-of-the-art BERT~\cite{devlin-etal-2019-bert} variant that produces more generalized word embeddings. 
Given this benefit, we input the question $q_x$ and dataset schema $s_x$ into RoBERTa, create their initial word embeddings $h_\text{initial}$, and then divide them into three parts
~Fig.~\ref{fig:rankingnetwork}(b). Next, 
a BiLSTM~\cite{hochreiter1997long}~(Fig.~\ref{fig:rankingnetwork}(c)) is employed to merge the embeddings of each data attribute name to have a~fused~representation~$v$.
Subsequently, a multi-head cross attention~(Fig.~\ref{fig:rankingnetwork}(d)) is introduced in the network to establish the connection between $v$ and $h_{q}$ and generate question-aware embedding $\widetilde{v}$~(Fig.~\ref{fig:rankingnetwork}(e)). Formally, the computation of the question-aware column embedding $\widetilde{v}_{c}$ is defined as:
\begin{align} 
{v}_{cj}^{q} &= MultiHeadAttn(v_{cj},~ h_{q},h_{q}, n)\\
\widetilde{v}_{cj} &= Norm~({v}_{cj}^{q} + {v}_{cj})
\end{align}
where $v_{cj}$ is the fused embedding of the column $j$, $n$ is the number of attention heads. For table embedding, we noticed that some questions only refer to columns, without mentioning the name of the table. Thus, we enhance ${v}_{t}$ with column information before computing the question-aware embedding~$\widetilde{v}_{t}$: 
\begin{align} 
{v}_{tk}^{c} &= MultiHeadAttn(v_{tk},~ v_{c}^{tk},v_{c}^{tk}, n)\\
{v}_{tk}^{q} &= MultiHeadAttn({v}_{tk}^{c},~ h_{q},h_{q}, n)\\
\widetilde{v}_{tk} &= Norm~({v}_{tk}^{q} + {v}_{tk}^{c})
\end{align}
where $v_{tk}$ is the fusion 
embedding of the table $k$, $v_{c}^{tk}$ refers to a set of fusion embeddings for the columns in table $k$. Finally, a binary classifier~(Fig.~\ref{fig:rankingnetwork}(f)) is appended to $\widetilde{v}$ to predict the probabilities that data attributes are relevant or irrelevant~to~the~question:
\begin{align} 
&v_i^* = ReLU~({W_{1} {\widetilde{v}_{i}} + {b_{1}}})\\
&P_{\text{re}_{i}}, P_{\text{irre}_{i}} = Softmax ({W_{2} v_i^* + {b_{2}}})
\label{eqn:relevance}
\end{align}
where ${W}$ and ${b}$ are trainable weight matrices and bias vectors of the classifier. $P_{\text{re}_{i}}$ and $P_{\text{irre}_{i}}$~(Fig.~\ref{fig:rankingnetwork}(g)) represent the possibility that the data attribute $i$ is relevant and irrelevant to the question, with the sum of both being equal to 1. The data attribute $i$ will be recognized as question-relevant if $P_{\text{re}_{i}}$ is greater than $P_{\text{irre}_{i}}$, and vice versa. In the end, we order the data attributes in the schema according to their $P_{\text{re}}$ value.

\textbf{Loss Function}~
To optimize the performance of $\mathcal{L}$ in identifying the links between datasets and questions, we trained the model with focal loss~\cite{mukhoti2020calibrating}.
It applies a modulating factor to down-weight the loss contribution of easy-to-classify samples during training and focus the model on hard-to-classify samples. The focal loss is calculated in the~following~form:

\begin{equation}
L =-\sum_{i=1}^{n} \alpha (1-p(t_{i}))^{\gamma} \log p(t_{i})
\end{equation}
where 
$p(t_{i})$ is the predicted probability that a data attribute is of the target class $t_i$~(i.e., question-relevant or question-irrelevant), $\gamma$ and $\alpha$ are a hyper-parameter, and the modulating factor $(1-p(t_{i}))^{\gamma}$ controls the loss contribution from samples. 

\textbf{Training Corpus}~~To train $\mathcal{L}$, we constructed a new corpus by linking the questions and datasets from an existing corpus introduced by Saparina \etal~\cite{saparina-osokin-2021-sparqling}. Specifically, the corpus in~\cite{saparina-osokin-2021-sparqling} was constructed by manually annotating the Spider dataset~\cite{yu-etal-2018-spider} based on the QDMR operations. 
It comprises 7,423 triplets of questions, datasets, and sequences of QDMR operations, in which the arguments of operations were annotated with corresponding data attributes. Using these annotations, we labeled data attributes as either relevant or irrelevant to questions, resulting in a new corpus with 7,423 pairs of questions and~labeled~datasets.

\textbf{Implementation}~~$\mathcal{L}$ was built upon \textit{RoBERTa-Large} and trained via 96 epochs with the gradient descent step of 4. The batch size and dropout rate were set to 8 and 0.2, respectively. The number of attention heads was set to 8. Adam~\cite{diederik2014adam} optimizer was used, and the learning rate was set to $10^{-5}$. $\gamma$ and $\alpha$ in the focal loss were 2.0 and 0.75. The training was conducted on an Ubuntu server with an NVIDIA~A100~80GB~GPU.

\vspace{-1em}
\subsection{Decomposition Network $\mathcal{D}$}
\label{sec:decomposition}

$\mathcal{D}$ is designed to generate a data analysis pipeline by taking a question and a ranked dataset schema as input, exploiting information from the question, the dataset, and the data--question links. To implement $\mathcal{D}$, we fine-tuned the pre-trained natural language model T-5~\cite{raffel2020exploring}, which is motivated by the following three reasons. First, T-5 is sensitive to the order of input sequences, so that the position information for data--question links can be effectively captured by it. Second, T-5 incorporates attention mechanisms that enable it to reason over the question and the ranked schema. Third, question decomposition can be recognized as a text2text task, in which T-5 has a strong~track~records~\cite{vaswani2018tensor2tensor,wang-etal-2020-rat}.


We present the conceptual calculation steps to give a brief idea about the model but leave the mathematical details to~\cite{ raffel2020exploring}. Generally, T-5 follows the encoder--decoder architecture. Given the question $q_x$ and ranked data schema $s_x'$, $\mathcal{D}$ first concatenates $q_x$ and $s_x'$ into a long text string and embeds it into an initial embedding $v$. Then, it encodes $v$ into a latent vector $h_e$ that captures the semantics of the data question $q$ and the corresponding data schema $s_x'$:
\begin{equation}
h_e = encode(v), ~~v = embed(q_x, s_x')
\end{equation}
Later, the decoder takes $h_e$ to compute a decoded latent vector $h_d$ and then uses it to compute the output probabilities of the tokens in the vocabulary given by the training samples via a softmax layer:
\begin{align} 
h_d &= decode(h_e)\\
P_{voc} &= Softmax(W h_d)
\end{align}
where $W$ is a weight matrix to be trained. In each round, the token in the vocabulary having the highest probability is chosen as the output of the model. In this way, the model generates a sequence of operations token by token in the following form:
\begin{equation}
\label{eq:decompose}
[op_1\{\mathtt{t}_{11},\ldots, \mathtt{t}_{1j}\}, \ldots,op_n\{\mathtt{t}_{n1}, \ldots, \mathtt{t}_{nk}\}] \leftarrow Decoder (h_e)
\end{equation}
where $\mathtt{t}_{ij}$ indicates the $j$-th token of the $i$-th operation.

\textbf{Loss Function}~~To encourage the output of $\mathcal{D}$ to be as identical as possible with the target analysis pipeline in our training corpus, the model was trained by using cross-entropy loss between the generated pipeline and ground truth:
\vspace{-1mm}
\begin{equation}
L =-\sum_{i=1}^{n} \log p(t_{i} \mid t_{1}, \ldots, t_{i-1}, q_x)
\label{eqn:decompositionloss}
\end{equation}
where $t_i$ is the current reference token in the target pipeline. Given the previous tokens $t_{1}, \ldots, t_{i-1}$ and the input data question $q_x$, this loss function tends to maximize the probability of the reference token $t_i$ as the prediction in the current step.

\textbf{Training Corpus}~~To train $\mathcal{D}$, we selected 5,349 records from the dataset constructed by Saparina \etal~\cite{saparina-osokin-2021-sparqling}, whose QDMR operations are indicated~in~Table~\ref{tab:operatorstable}. 

\textbf{Implementation}~~$\mathcal{D}$ was implemented by fine-tuning the \textit{Flan-T5-large} via 96 epochs with the gradient accumulation step as 16. The batch size was set to 8. Adam~\cite{diederik2014adam} optimizer was used and the learning rate was set to $10^{-4}$. The beam search decoding method was employed with the beam size set to 5. The training was conducted on an Ubuntu server with an NVIDIA A100 80GB GPU.

\vspace{-1em}
\section{Damation Generation and Interface}
In this section, we first introduce action-oriented unit visualization and how we use it to represent data analysis pipelines as datamations. After that, we describe the design of the natural language interface and the corresponding interactions.
\vspace{-2em}
\subsection{Action-Oriented Unit Visualization}
\label{section:low-level}
Data analysis pipelines enable the flow of data from the source to the answer. To visualize them in an easy-to-understand manner, the presentations of data analysis pipelines should fulfill three requirements. First, the presentations should show data flow in a coherent manner~(\textbf{DR1}). Second, the presentations are supposed to be detailed enough to illustrate how data has been manipulated in each step~(\textbf{DR2}). Third, the presentations should explicitly explain the meaning of analysis operations~(\textbf{DR3}). For example, when showing the operation \texttt{FILTER}, the presentations should specify the filtering criteria so that users can understand the operation.

To fulfill these requirements, we designed an animated action-oriented unit visualization to illustrate data analysis pipelines in the form of datamation~\cite{pu2021datamations}. In particular, Pu~\etal~\cite{pu2021datamations} have shown that the datamation is effective in helping people understand the process involved in analyzing data. Following this idea, we present data analysis pipelines as datamations, animating the analysis operations within the pipeline and their corresponding data changes~(\textbf{DR1}). Moreover, to ensure that the datamation is flexible enough to illustrate how data has been manipulated across the pipeline, we use unit visualization~\cite{drucker2015unifying}, which represents each data item by an individual mark, to help readers track the trajectory of data change easily~(\textbf{DR2}). We created a set of low-level actions~(introduced in the next paragraph) to control the visual channels, appearance, and layout of unit visualization, allowing us to manipulate the visualization dynamically and present the data analysis pipeline as a datamation. Additionally, to explain the meaning of operations, we add annotations and captions to the datamation to communicate meaning~(\textbf{DR3}).

We developed a chart library based on \textbf{D3.js}~\cite{bostock2011d3} to implement our action-oriented unit visualization. The library takes a sequence of low-level actions as input and renders a datamation as output. These actions are designed to illustrate QDMR operations and fell into the following~three~types:

\textbf{Data Actions.}~~The data actions are designed to update data items represented in the unit visualization, which includes four actions used to (1)~\textbf{\textit{select}} a set of qualified data items; (2)~\textbf{\textit{filter}} the current data items based on a condition; 
 (3)~\textbf{\textit{aggregate}} the data items to calculate a statistic measure such as minimum, mean, and sum; or (4)~\textbf{\textit{sort}} the data items in an order. When these actions are performed, the underlying data will be manipulated, and the visualization layout will be updated accordingly to add, remove, or rearrange the units shown in the~visualization.

\begin{figure}[!tbh]
\setlength{\abovecaptionskip}{10pt}
\centering 
\vspace{-0.5em}
\includegraphics[width=0.9\columnwidth]{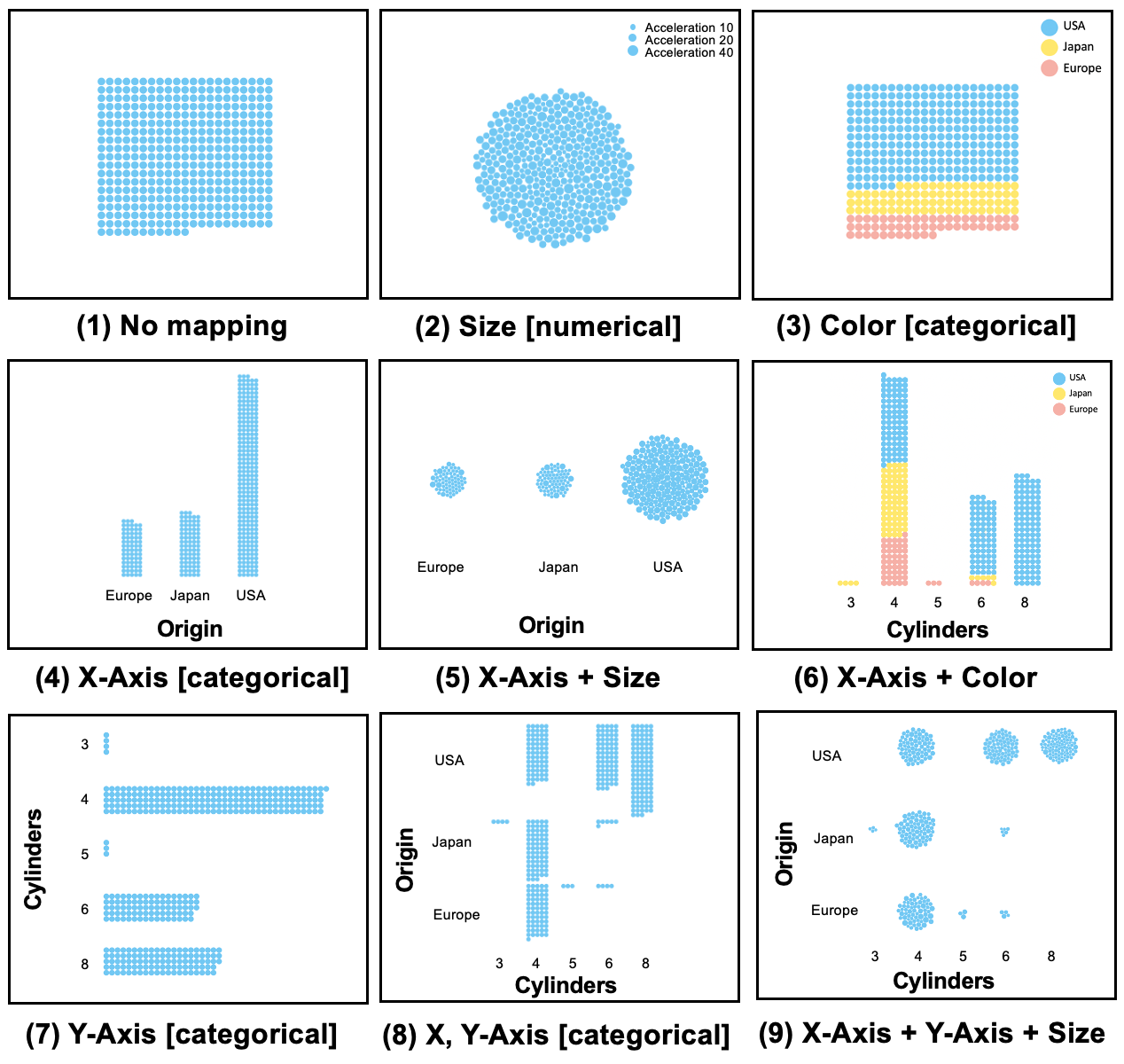}
\vspace{-1em}
\caption{The appearance of the unit visualization is controlled by a set of visualization actions.}
\vspace{-1em}
\label{fig:unitvisexample}
\end{figure}

\textbf{Visualization Actions.}~~The visualization actions are utilized to control the data mapping of visual channels. In our design, the visual appearance (position, color, and size) of each unit in the visualization is determined by the data mappings through four independent visual channels, i.e., x-axis, y-axis, unit size, and unit color, designed for mapping different types of data attributes. In precise, \textbf{\textit{x}} and \textbf{\textit{y}} actions are used for mapping categorical or temporal data attributes via \{x\}-axis and \{y\}-axis, resulting in group views for the data with discrete temporal or categorical attributes (Fig.~\ref{fig:unitvisexample}(4--9)). The \textbf{\textit{size}} action is designed to encode a numerical data attribute by the size of a unit. By default, 
the units in the visualization are packed in a square form to facilitate counting (Fig.~\ref{fig:unitvisexample}(1)). When the size channel is used to map a numerical attribute, the size of each unit is proportional to the corresponding attribute value, and the layout of the units will be changed to circle packing to avoid the overlaps of the units with different sizes (Fig.~\ref{fig:unitvisexample}(2)). 
Finally, the \textbf{\textit{color}} action is used to map a categorical data attribute by the filling colors~(hue) of the units (Fig.~\ref{fig:unitvisexample}(3,6)). Combining these actions will provide flexible data mapping strategies, resulting in a variety of unit visualization forms with different layouts, as shown in Fig.~\ref{fig:unitvisexample}.

\begin{figure}[!tbh]
\setlength{\abovecaptionskip}{10pt}
\centering 
\vspace{-1em}
\includegraphics[width=0.9\columnwidth]{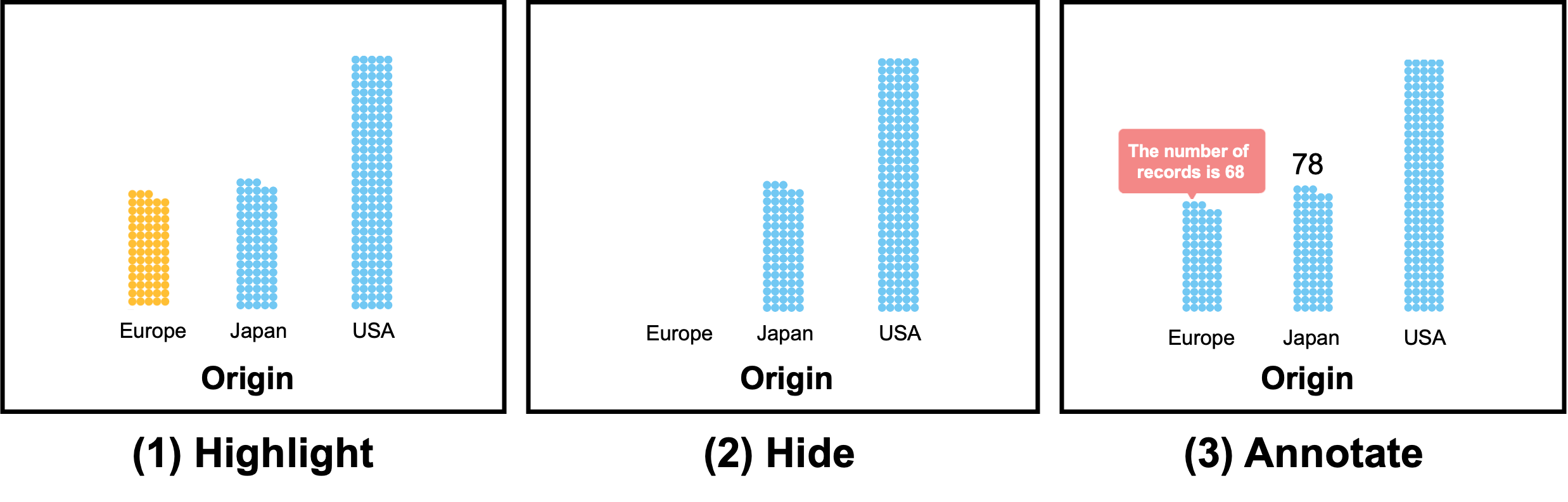}
\vspace{-1em}
\caption{The appearance of the unit visualization is controlled by a set of annotation actions.}
\vspace{-0.75em}
\label{fig:annotateexample}
\end{figure}

\textbf{Annotation Actions.}~~The annotation actions are used to emphasize the information within the visualization. We designed three types of annotation actions to (1)~\textbf{\textit{highlight}} a focal unit by changing its filling color~(Fig.~\ref{fig:annotateexample}(1)); (2)~\textbf{\textit{hide}} the non-focal units by making it invisible, i.e., hidden in the view~(Fig.~\ref{fig:annotateexample}(2)); (3)~\textbf{\textit{annotate}} a unit or a group of units by adding a textual tooltip or a text that shows the customized information~(Fig.~\ref{fig:annotateexample}(3)). 

To composes these low-level actions together, we designed animated transitions between them, following principles and empirical findings in visualization literature. First of all, transitions only contain necessary and meaningful motion~\cite{fisher2010animation}, as each animated transition is directly linked to an action in the sequence. Furthermore, transitions are in a slow-in/slow-out manner to facilitate viewer understanding when animating between views~\cite{dragicevic2011temporal}. Moreover, when data is grouped, all groups move together rather than one after the other, since there is little advantage to ``staggering" group animations~\cite{chevalier2014not}. Likewise, data points within a group move in the same way because things with similar motion are perceived as belonging to the same group~\cite{chalbi2019common}. Last but not least, we present motions step by step using staged~animation to guide users' attention and facilitate data perception, as suggested~in~\cite{heer2007animated}.

\vspace{-1em}
\subsection{Datamation Generation}
Based on the action-oriented unit visualization, we generate datamations via the following three steps:

\textbf{Validating.}~First of all, we validate data analysis pipelines generated by our algorithm and use the valid one to generate datamation. Specifically, when resolving a question, we employ the beam search decoding method~\cite{sutskever2014sequence} to get a set of candidate pipelines. In a nutshell, beam search is an inference strategy in neural sequence models. It maintains the $n$~top-scoring partial sequences and expands them in a greedy left-to-right fashion. 
The candidate pipelines are verified for correct syntax by a set of rules that we summarized based on QDMR definitions. For example, the operation \texttt{PROJECT} requires one argument that should refer to a column in the dataset. A data analysis pipeline that violates this rule will be considered invalid. In this way, we verify candidate pipelines one by one and choose the first valid pipeline for the next generation step.


\textbf{Translating.}~In the second step, we sequentially translate the operations in the data analysis pipeline to low-level actions. In precise, we tailor a set of low-level actions for each type of analysis operation to facilitate interpretation~(details are shown in Table~\ref{tab:operatorstable}). Once an analysis operation is translated to low-level actions, these actions are executed in the pre-defined order, where data actions are first performed to update the data items shown on the view, and then visualization actions and annotation actions gradually modify encoding methods and annotations in the follow-up steps. For example, the \texttt{FILTER} operation maps to the data action \textit{filter}, which is performed to choose a subset of data. After that, the annotation action \textit{highlight} is used to draw viewers' attention to these data items, and the annotation action \textit{hide} is used to remove the data items that do not meet the condition from~the~visualization.

\textbf{Captioning.}~Third, to help communicate the operation's meaning and effects, we generate a caption for each operation based on a pre-defined natural language template. The caption is designed to provide context information~(e.g., a filter condition) and describes the changes in the data and the visual mappings. It is displayed on top of each view and updated gradually during the animation process. For example,  the \texttt{AGGREGATE} operation is described as ``the maximum/minimum/average/total value of [a numerical attribute] of the following [units] is [a numeric value]".

We use the data analysis pipeline introduced in Section~\ref{sec:qdmr} as an example to illustrate how to generate a datamation. The input data analysis pipeline is:
\[
[\texttt{SELECT}][\texttt{PROJECT}][\texttt{FILTER}][\texttt{AGGREGATE}] 
\]
which can be translated into an action sequence as follows:
\[
[select,layout][x-axis][filter,highlight]
\]
\[
[hide][aggregate,annotate]
\]
Here, we leave out the arguments in analysis operations and low-level actions for easy reading.
Semantically, it indicates to (1)~\textit{select} a collection of data records, i.e., students, from the data source and represents each record as a unit and layout all the records in the unit visualization; (2)~\textit{encode} the attribute, i.e., year of birth,  by x-axis; (3)~\textit{filter} the records that satisfy the condition and  \textit{highlight} them by highlighting each qualified unit color and (4)~\textit{hide} those unqualified ones; (5)~\textit{aggregate} the records, i.e., students born in 2000, by counting the units and  \textit{annotate} on records via a tooltip to illustrate the total number of students. Finally, captions are generated based on templates and added to datamation to explain the operation's~meaning~and~effects. 

\vspace{-1em}
\subsection{Natural Language Interface and Interactions}

\begin{figure}[!tbh]
\setlength{\abovecaptionskip}{10pt}
\centering 
\includegraphics[width=\columnwidth]{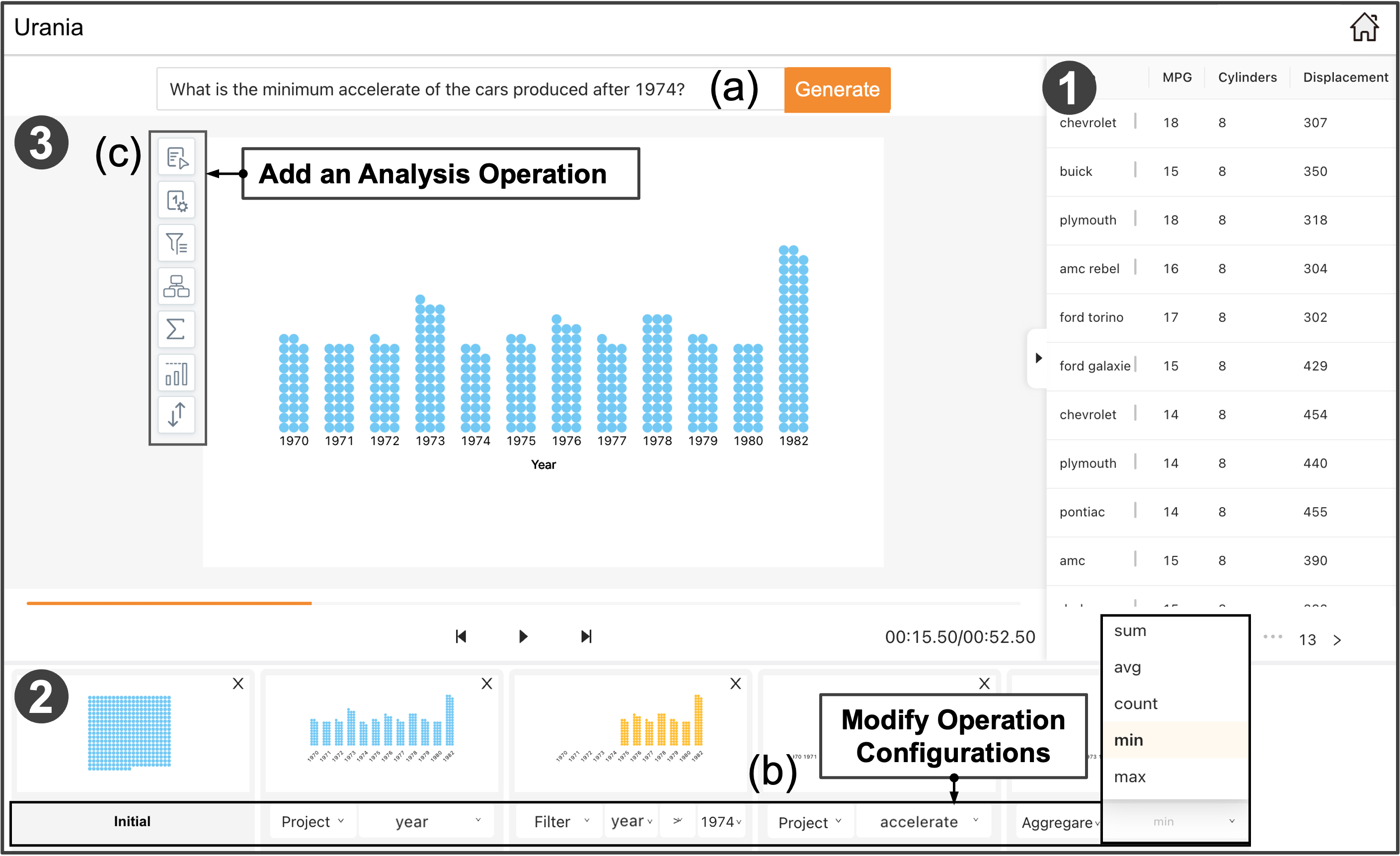}
\vspace{-1.5em}
\caption{The natural language interface consists of three major views: (1) the data view for previewing uploaded data; (2) the key-frame view for illustrating and editing the analysis operations; (3) the datamation view for showing data analysis pipelines and adding new analysis operations.}
\label{fig:userinterface}
\vspace{-1em}
\end{figure}

The natural language interface of \name, as shown in Fig.~\ref{fig:userinterface}, consists of three major views: the
\textit{data view}~(Fig.~\ref{fig:userinterface}-1), the \textit{key-frame view}~(Fig.~\ref{fig:userinterface}-2), and \textit{datamation view}~(Fig.~\ref{fig:userinterface}-3). The \textit{data view} is designed to illustrate the uploaded data, enabling users to easily access the data during the analysis process. Users can preview the data and enter a question of interest into the input box~(Fig.~\ref{fig:userinterface}-(a)). To assist with the ``cold start'' problem in the initial stages of data exploration, several analytic questions are recommended in the input box to help users~begin~their~exploration. Once a question is entered, it is sent to the question decomposition algorithm to generate a data analysis pipeline that resolves the question.

The \textit{key-frame view}~(Fig.~\ref{fig:userinterface}-2) illustrates the analysis operations in the generated pipeline and allows users to edit them. Specifically, the operations and their corresponding unit visualization key-frames are presented in sequence to help users understand the intermediate result of each operation. In addition, the drop-down menus under key-frames, as shown in Fig.~\ref{fig:userinterface}-2(b), provide users with access to the operation's parameters, allowing them to refine or expand the operation for a second round of exploration. 

The \textit{datamation view} (Fig.~\ref{fig:userinterface}-3) visualizes the data analysis pipeline as a datamation and enables users to add new analysis operations to the pipeline. In this view, the datamation is played as a video, where users can pause or replay it via playback buttons at the bottom. Moreover, users can use shortcut buttons~(Fig.~\ref{fig:userinterface}-3(c)) to 
add new operators to the existing data analysis pipeline. Specifically, we created seven shortcut buttons, one for each type of analysis operation, to add new operations to the end of the existing pipeline. Clicking the shortcut button will bring up a pop-up window that helps users set up an analysis operation in a few seconds. It is worth mentioning that users are allowed to create a data analysis pipeline from scratch using the shortcut buttons without inputting a data question. Once existing operations are modified or new operations are added, the key-frames and the datamation shown in the interface will be updated accordingly. 

\section{Evaluation}

We evaluated the data-aware question decomposition algorithm via two quantitative experiments, and assessed the usability of \name through interviews with experts.
\vspace{-1em}
\subsection{Quantitative Evaluation}
To demonstrate the performance of our algorithm, we conducted two experiments, one on the correctness of $\mathcal{L}$, and another on the effectiveness of $\mathcal{L}$ and the accuracy of our algorithm $\mathcal{L+D}$.

\textbf{Correctness.}~We measured the correctness of $\mathcal{L}$ for classifying data attributes as question-relevant or question-irrelevant. In this experiment, we used 80\% of the corpus~(introduced in Section~\ref{sec:ranking}) as the training set and another 20\% of the corpus as the testing set. 

The performance of $\mathcal{L}$ in the testing set is summarized in Table~\ref{tab:rankingevaluation}. We can see that $\mathcal{L}$ can precisely classify the data attributes, achieving a classification F1 score of 97.6\% and 94.2\% for tables and columns, respectively. With this exceptional performance, we confirm that the data--question links identified by $\mathcal{L}$ are valid, and the ranked schema could provide useful knowledge for $\mathcal{D}$ to learn the conversion between inputs and outputs.
\vspace{-1em}
\begin{table}[htb]
\caption{The performance of \modelname on the testing set.}
\setlength\aboverulesep{0pt}
\setlength\belowrulesep{0pt}
\renewcommand{\arraystretch}{1.3}
\centering
\def\arraystretch{1.3}
\begin{tabular}{cccc}
\toprule
\rotatebox{0}{Type} & \rotatebox{0}{Precision} & \rotatebox{0}{Recall} & \rotatebox{0}{F1-score}
\\\midrule
    Table & 0.968 & 0.985 & 0.976 \\
Column &  0.929 & 0.957 & 0.942

\\\bottomrule
\end{tabular}
\label{tab:rankingevaluation}
\vspace{-0.75em}
\end{table}

\textbf{Effectiveness and Accuracy.}~To evaluate the effectiveness of $\mathcal{L}$ and the accuracy of our algorithm $\mathcal{L+D}$, we compared the performance of question decomposition models with and without the help of $\mathcal{L}$, while analyzing the performance difference between $\mathcal{L+D}$ and several baseline models. We used execution accuracy as the metric to evaluate these models. It compares the execution results of generated pipelines with target pipelines. To be able to execute them, we adopted a QDMR-to-SQL program~\cite{wolfson-etal-2022-weakly} to convert both generated pipelines and target pipelines to executable SQL queries. 

In the experiment, we selected three baseline decomposition models as references. The first baseline model ($\mathcal{B}_1$) is a sequence-to-sequence neural network with a 5-layer LSTM encoder and cross-attention. The second one ($\mathcal{B}_2$) is another LSTM-based sequence-to-sequence model that incorporates the copy mechanism~\cite{gu2016incorporating} to deal with unseen questions. 
The third baseline model ($\mathcal{B}_3$) is the state-of-the-art technique proposed by Saparina and Osokin~\cite{saparina-osokin-2021-sparqling}, built on the relation-aware transformer~\cite{wang-etal-2020-rat} and~grammar-based~decoder~\cite{yin-neubig-2017-syntactic}.

We trained two versions of $\mathcal{B}_1$, $\mathcal{B}_2$ and $\mathcal{D}$ based on the corpus introduced in Section~\ref{sec:decomposition}; one was trained using the unordered dataset schema, and another was trained using the ranked dataset schema from $\mathcal{L}$. All models were trained on 80\% of the corpus and tested on another 20\% of the corpus. To evaluate the ability of models to generalize across different underlying datasets, we carefully picked the training samples and testing samples so that no two identical datasets would be in both the training samples and testing samples. Moreover, to better understand models' performance on different questions, we used the question hardness criteria~\cite{yu-etal-2018-spider} to divide the questions into four levels of complexity, i.e., easy, medium, hard, and extra-hard.

The evaluation results on the testing set are summarized in Table~\ref{tab:decompositionevaluation}. With the help of $\mathcal{L}$, the performance of $\mathcal{B}_1$, $\mathcal{B}_2$, and $\mathcal{D}$ has improved on different hardness levels of questions, the overall execution accuracy of them has increased by 10 to 35 percent. This demonstrates that $\mathcal{L}$ is effective, the data--question links in the ranked schema can help various decomposition algorithms generate more accurate results. Moreover, our algorithm~$ \mathcal{L+D} $ performs well on different hardness levels of questions. It achieves 80.0\% on easy questions, 72.2\% on medium questions, 71.0\% on hard questions, and 61.7\% on extra-hard questions, outperforming baseline models at all four levels. In summary, the data-aware question decomposition algorithm shows its effectiveness by achieving an overall execution accuracy of 73.4\%, which is higher than the state-of-the-art technique~$\mathcal{B}_3$.


\begin{table}[htb]
\vspace{-0.5em}
\caption{Comparative results for question decomposition based on execution accuracy.}
\setlength\aboverulesep{0pt}
\setlength\belowrulesep{0pt}
\renewcommand{\arraystretch}{1.3}
\centering
\def\arraystretch{1.3}
\begin{tabular}{cccccc}
\toprule
\rotatebox{0}{Models} & \rotatebox{0}{Easy}  & \rotatebox{0}{Medium}  & \rotatebox{0}{Hard} & \rotatebox{0}{Extra-Hard} & \rotatebox{0}{Overall} 
\\\midrule
    $\mathcal{B}_1$ & 28.2\% & 11.8\% & 4.7\% & 3.4\% & 15.5\% \\
$\mathcal{B}_2$ & 32.6\% & 13.8\% & 12.7\% & 3.4\% &18.8\%  \\
$\mathcal{L}+\mathcal{B}_1$ & 41.8\% & 20.9\% & 15.8\% & 6.8\% &25.8\% \\
$\mathcal{L}+\mathcal{B}_2$ & 40.7\% & 21.7\% & 23.8\% & 10.3\% & 27.0 \%  \\
$\mathcal{D}$ & 44.0\% & 36.7\% & 43.4\% & 36.1\% & 39.8\%  \\
$\mathcal{B}_3$ & 64.3\% & 60.4\% & 57.34\% & 48.3\% &60.1\%  \\
\textbf{{$\mathcal{L}+\mathcal{D}$}} & \textbf{{80.0\%}} & \textbf{{72.2\% }} & \textbf{{71.0\%}} & \textbf{{61.7\%}} & \textbf{{73.4\%}} 

\\\bottomrule
\end{tabular}
\label{tab:decompositionevaluation}
\vspace{-2.0em}
\end{table}


\subsection{Expert Interviews}
To assess Urania's usability, we held a semistructured interview with three domain experts, represented as \textbf{E1}, \textbf{E2}, and \textbf{E3}. In particular, \textbf{E1} is an industry analyst who has worked in a consulting company for three years, focusing on analyzing industry data using tools like PowerBI~\cite{powerbi} or Tableau~\cite{tableau} to explore business opportunities. \textbf{E2}, a software developer, has four years of experience in developing data analysis and exploration systems. \textbf{E3} holds the position of CEO in a company that develops visual analytics dashboards.

\textbf{Datasets.} We selected three datasets for the case study and had been respectively distributed them to the experts. The first dataset, \textbf{D1}, records information about 168 flights, it contains the flight number, date, number of passengers, flight distance, departure country, and four other dimensions~(2 nominal, 1 geographic, and 1 quantitative). The second dataset \textbf{D2} contains the personal information of 136 graduates from a university. It includes graduates' education level, employer, salary, city, graduation year, and four other dimensions~(2 nominal and 2 quantitative). The third dataset \textbf{D3} is related to 389 vehicles, including their name, weight, horsepower, production country, production year, and four other dimensions~(1 nominal and 3 quantitative). We chose these datasets because they are of real-world interest, have similar complexity, and feature a reasonable amount of data to ensure that a comprehensive understanding could be obtained within the study.

\textbf{Procedure and Task.} All interviews were conducted through one-on-one offline meetings. At the beginning of every interview, we explained the purpose of our study and introduced the dataset that will be used. We provided two natural language systems, \name and Tableau Ask Data~\cite{tableau}, to experts for their hands-on experience. We spent about 15 minutes giving a brief demonstration of \name and Ask Data. Following that, each expert was invited to use both systems to explore the given dataset via natural language queries. We encouraged all the experts to verbally summarize their findings and thoughts in a think-aloud protocol. Experts were allowed to take as much time as they needed to conduct their exploration. Each session took around an hour, we saved the questions asked by experts and the answers were returned by systems. After exploring the data, experts were invited to have an interview in which we collected their comments about two systems on three aspects: (1) the quality of the answers in terms of interpretability and reliability; 
(2) the interactions, and (3) the system's usability in EDA. Each interview lasted for about half an hour with the processes recorded.

\textbf{Baseline System.} Tableau Ask Data~\cite{tableau} is the state-of-the-art natural language-based analytic commercial system. It answers a user's input question via a chart or a direct result such as a number. We chose this system as the baseline for comparison for two reasons. First, both Ask Data and Urania are designed for data exploration, not visualization authoring. Second, other relevant NLIs~\cite{liu2021advisor, shi2021talk2data} are not open source or released to the public, so they cannot be compared in the experiment.

\textbf{Study Result.} 
We reviewed the answers provided by both systems during the exploration and then summarized the comments from experts that were collected during~the~interview.

\begin{figure}[!tbh]
\setlength{\abovecaptionskip}{10pt}
\centering 
\vspace{-1em}
\includegraphics[width=0.9\columnwidth]{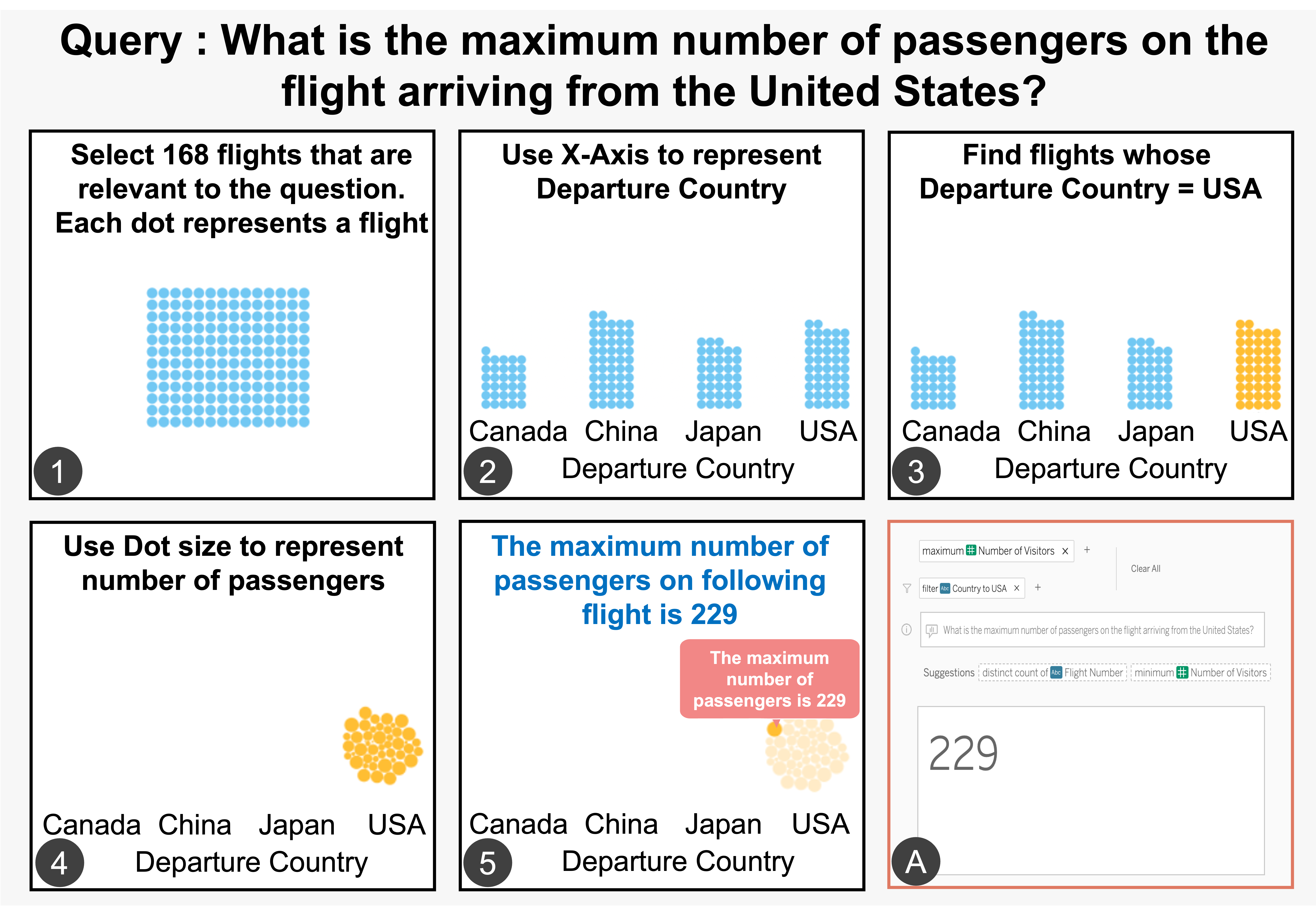}
\vspace{-1em}
\caption{Answers from \name~(1-5) and Ask Data~(A) to an \textbf{E1}'s question.}
\label{fig:casestudy1}
\vspace{-0.5em}
\end{figure}

\begin{figure}[!tbh]
\setlength{\abovecaptionskip}{10pt}
\centering 
\vspace{-1em}
\includegraphics[width=0.9\columnwidth]{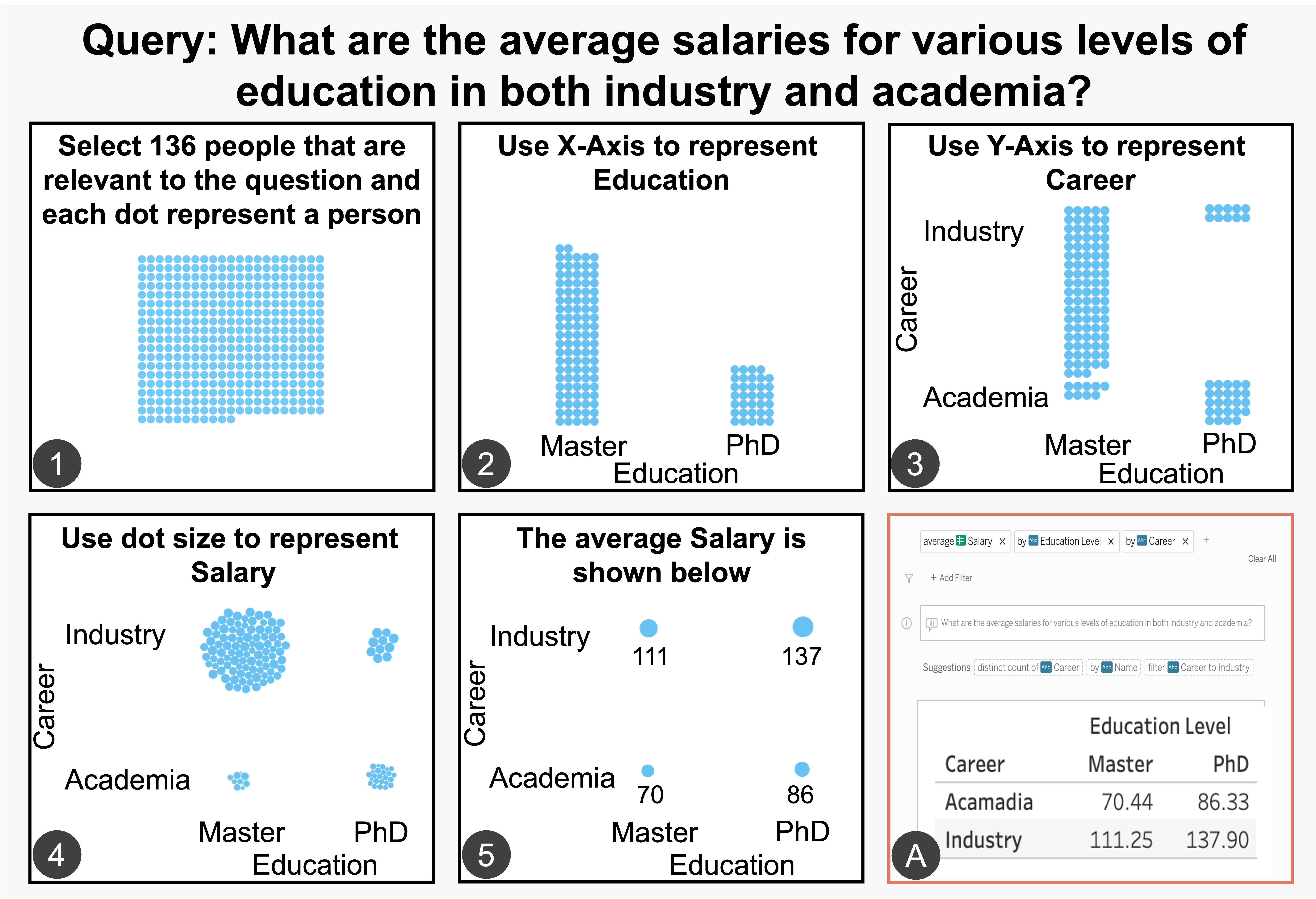}
\vspace{-1em}
\caption{Answers from \name~(1-5) and Ask Data~(A) to an \textbf{E2}'s question.}
\label{fig:casestudy2}
\end{figure}

\begin{figure}[t]
\setlength{\abovecaptionskip}{10pt}
\centering 
\vspace{-1em}
\includegraphics[width=0.9\columnwidth]{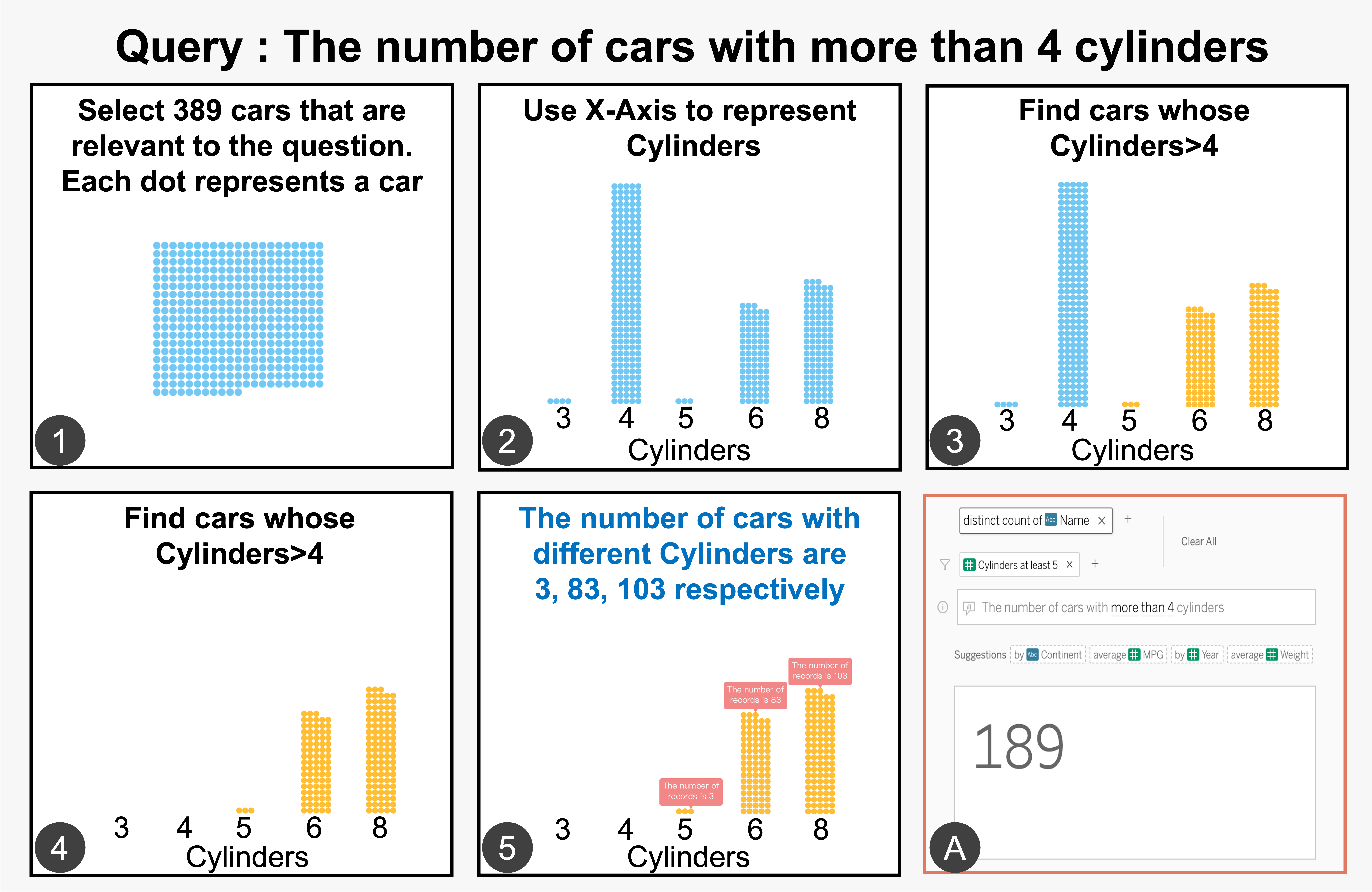}
\vspace{-1em}
\caption{Answers from \name~(1-5) and Ask Data~(A) to an \textbf{E3}'s question. }
\label{fig:casestudy3}
\vspace{-1.5em}
\end{figure}

\underline{\textit{Example Cases.}} We select three questions, one from each expert, that were correctly resolved by both systems and illustrate their answers in Fig.~\ref{fig:casestudy1}, Fig.~\ref{fig:casestudy2}, and Fig.~\ref{fig:casestudy3}. In these figures, the answers from \name are marked in black, whereas the answer provided by Ask Data are marked in red.

Fig.~\ref{fig:casestudy1} shows the answers regarding the query: ``what is the maximum number of passengers on the flight arriving from the United States?", asked by \textbf{E1} based on the flight dataset (\textbf{D1}). The key-frames of the datamation generated by \name are shown in Fig.~\ref{fig:casestudy1}(1-5). To answer this question, \name first grouped the records by countries~(\textit{Frame} 2) and then narrowed down to flights from the USA~(\textit{Frame} 3). Eventually, it obtained the number of passengers on each flight~(\textit{Frame} 4) and calculated the maximum value~(\textit{Frame} 5). On the other hand, the answer provided by Ask Data is shown in Fig.~\ref{fig:casestudy1}(A). It directly returned the calculated result, 229. 

Fig.~\ref{fig:casestudy2} shows the responses related to the question: ``what are the average salaries for various levels of education in both industry and academia?", asked by \textbf{E2} based on the graduate dataset (\textbf{D2}). \name first grouped the records by education~(\textit{Frame} 2) and career~(\textit{Frame} 3). After that, it retrieved the salary of each graduate~(\textit{Frame} 4) and calculated the average value in each group~(\textit{Frame} 5). In addition, Ask Data provided a list of average salaries in different groups. This data can be visualized using either a bar chart, a heatmap, or a table view. To present the results clearly within the limited figure space, we opted to use the table view as an example.

\underline{\textit{Interview Feedback.}} We received a number of valuable comments from experts that were summarized below:

\textit{The quality of answers.} From the aspect of interpretability and reliability, experts unanimously agreed that the datamation generated by \name is more convincing~(\textbf{E1}, \textbf{E3}) and safer~(\textbf{E2}) than the static answers~(i.e., numbers or charts) from Ask Data. \textbf{E1} commented that ``showing the analysis process via an animation is intuitive ... it [datamation] is better than the answer returned by another system [Ask Data] as I can easily tell if the answer is correct". \textbf{E2} appreciated our design and commented that ``it is important for an automatic EDA system to protect the user from false-discoveries, ..., datamation is a promising solution to decrease the chance of users being misled". \textbf{E3} commented that ``it [datamation] helps me understand how the answer was derived step by step and pick up some useful insights".

\textit{The interactions in systems.} All the experts were satisfied with exploring the data via natural language questions. \textbf{E1} mentioned that ``two systems [Urania and Ask Data] are ``hands-free", allowing for an easy exploration". \textbf{E2} and \textbf{E3} both agreed that the interactivity of the generated results is useful. In particular, \textbf{E2} identified that  `` in  the first system [Urania], I don't need to enter a new query over and over again ... does not disrupt my exploration". \textbf{E3} commented that ``the interactions [in Urania] are able to make the auto-generated sessions personalized ... I could feel I was still in the ``driver’s seat'' and in control of the exploration process".

\textit{System usability.}
All experts thought that \name is  more useful and helpful in EDA compared with Ask Data. \textbf{E1} emphasized the importance of the exploration process and commented that ``it [the exploration process] is more valuable to me than an answer, especially for a new dataset". \textbf{E2} identified that ``I think the first system [Urania] is the better choice for data exploration because it includes an explanation for each generated exploratory step, describing the results and surfacing potential insights". \textbf{E3} noted that \name ``strikes a balance between manual and automatic exploration, making it a valuable solution for personalizing and making auto-generated exploratory sessions interactive". Overall, the experts' feedback suggests that \name can enhance the exploration process and lead to safe and valuable insights.

\textbf{Analysis.} According to experts' feedback, we analyzed the technical differences that theoretically make \name perform better than Ask Data. First, a major difference between \name and Ask Data is that we use datamtions to present how answers are uncovered step by step. Thus, the answers from \name are more reliable and provide additional context information about underlying data. Second, the interactive design of \name enables users to continuously examine the data in a more efficient and intuitive manner. This is particularly useful when users are inspired by existing answers and want to delve deeper into the data. Third, \name uses a deep neural network to understand the relationships between the question and the given dataset at a semantic level, rather than relying on simple character matching. This allows \name to effectively recognize data attributes that are mentioned explicitly (i.e., refer to attribute names) and implicitly (i.e., refer to an attribute's alias) in the question, making \name a more effective solution than Ask Data for handling vague or~under-specified~questions.


\section{Discussion, Limitations, and Future Work}

By using \name in EDA, users can get datamations as the answers to their questions and thus explore a new dataset more efficiently. Still, our system has several limitations that were found during the implementation or mentioned by experts. We discuss these limitations and potential future work below.

\paragraph{ \bf Extending the variety of operations.}

\name currently supports seven types of operations that are frequently used in the analysis process~\cite{amar2005low}. However, there is a limited range of analytical tasks that these actions can perform. For instance, we found that experts asked ``true or false" queries in the experiment, which our system cannot answer due to the absence of relevant operations. Enhancing the variety of operations is crucial to implement our system in a~real-world~setting.


\paragraph{\bf Optimizing datamation generation.} To create datamations, we have pre-defined a set of low-level actions for each type of analysis operation. These low-level actions are tailored to the semantic meaning of the operation to facilitate the understanding of users. However, the characteristics of the data are also an important feature that should be considered when illustrating the operations. For example, as shown in Table~\ref{tab:operatorstable}, the \texttt{PROJECT} operation is mapped to a visual action that encodes the specified column in a visual channel.  
A categorical/nominal column can be represented using the x-axis, y-axis, or unit color. However, if the categories in that column are diverse, encoding the column with the unit color might be less effective~\cite{mackinlay1986automating}. Therefore, an intelligent approach is needed in future work to optimize the datamation generation process. It should be able to select appropriate low-level actions in different conditions and optimize the duration of each action.

\paragraph{ \bf Incorporating with large language models.} 
With the rapid development of language models, GPT-3.5~\cite{radford2018improving} is able to accomplish many language-related tasks, such as paraphrasing sentences, answering open-domain questions, writing computer programs, etc. Although GPT-3.5 is versatile, at the time we wrote this paper, it is paid to use and has some restrictions on the regions in which it can be used. Still, we believe that the open-sourced language model outperforming our current algorithm will be proposed in the near future. Accordingly, we need to keep enhancing our algorithm by incorporating more advanced language models. 
However, large language models are usually designed for general-purpose. To utilize them in our task, we have to either prepare sufficient training data or adopt data-efficient strategies to fine-tune the models with limited data.
\vspace{-1em}
\section{Conclusion}
In this paper, we present \name, an interactive system visualizing the data analysis pipelines that answer input questions. Given a dataset and a natural language question, \name can identify the necessary steps to answer the question and present them in the form of a datamation. Quantitative experiments have proven the effectiveness of the question decomposition algorithm, and expert interviews show that \name is useful and helpful for EDA.
\vspace{-1em}
\section*{ACKNOWLEDGMENTS}
Nan Cao is the corresponding author. This work was supported in part by NSFC 62061136003, NSFC 6200070909, NSFC 62072338, and NSF Shanghai 23ZR1464700.

\bibliographystyle{IEEEtran}
\bibliography{main}

\vspace{12pt}

\end{document}